\documentstyle[12pt]{article}  

\def\thefootnote{\fnsymbol{footnote}}
\def\ref#1{$^{#1)}$}

\newcommand{\DaDa}{{\cal D}^{\alpha}{\cal D}_{\alpha}}
\newcommand{\DbDb}{{\cal D}_{\dot{\alpha}}{\cal D}^{\dot{\alpha}}}

\def\[{\left [}
\def\]{\right ]}
\def\({\left (}
\def\){\right )}

\def\CO{{\cal O}}

\begin{document}
\begin{titlepage}
\begin{center}
            \hfill    LBNL-40617 \\
%            \hfill    UCB-PTH-97/xx \\
            \hfill    hep-ph/9707498 \\[0.03in]
{\large \bf Recent Progress in Weakly-Coupled \\ 
            Heterotic String Phenomenology}
\footnote{This work was supported in part by the Director, Office of 
Energy Research, Office of High Energy and Nuclear Physics, Division 
of High Energy Physics of the U.S. Department of Energy under 
Contract DE-AC03-76SF00098 and in part by the National Science 
Foundation under grant PHY-95-14797.}\\[.1in]

                 Yi-Yen Wu \\[.05in]

{\em  Theoretical Physics Group \\
      Ernest Orlando Lawrence Berkeley National Laboratory \\
      University of California, Berkeley, California 94720 \\
      and \\
      Department of Physics \\
      University of California, Berkeley, California 94720}\\[.1in]
\end{center}

\begin{abstract}
Some recent developments in the weakly-coupled heterotic string 
phenomenology are reviewed. We discuss several important issues
such as dilaton/moduli stabilization, supersymmetry breaking (by 
hidden-sector gaugino condensation), gauge coupling unification 
(or the Newton's constant), the QCD axion, as well as 
cosmological problems involving the dilaton/moduli and the axion.
(Talk given at the 5th International Conference on Supersymmetries
in Physics, May 27-31, 1997, Philadelphia, PA, USA)
\end{abstract}
\end{titlepage}
\renewcommand{\thepage}{\roman{page}}
\setcounter{page}{2}
\mbox{ }

\vskip 1in

\begin{center}
{\bf Disclaimer}
\end{center}

\vskip .2in

\begin{scriptsize}
\begin{quotation}
This document was prepared as an account of work sponsored by the United
States Government.  Neither the United States Government nor any agency
thereof, nor The Regents of the University of California, nor any of their
employees, makes any warranty, express or implied, or assumes any legal
liability or responsibility for the accuracy, completeness, or usefulness
of any information, apparatus, product, or process disclosed, or represents
that its use would not infringe privately owned rights.  Reference herein
to any specific commercial products process, or service by its trade name,
trademark, manufacturer, or otherwise, does not necessarily constitute or
imply its endorsement, recommendation, or favoring by the United States
Government or any agency thereof, or The Regents of the University of
California.  The views and opinions of authors expressed herein do not
necessarily state or reflect those of the United States Government or any
agency thereof of The Regents of the University of California and shall
not be used for advertising or product endorsement purposes.
\end{quotation}
\end{scriptsize}

\vskip 2in

\begin{center}
\begin{small}
{\it Lawrence Berkeley Laboratory is an equal opportunity employer.}
\end{small}
\end{center}

\newpage
\renewcommand{\thepage}{\arabic{page}}
\def\thefootnote{\arabic{footnote}}
\setcounter{page}{1}
\setcounter{footnote}{0}
%THIS IS PAGE 1 (INSERT TEXT OF REPORT HERE)
\section{Introduction}
Superstring theory is known to offer a very powerful 
scheme of supersymmetry phenomenology, and in the past the heterotic string 
theory was the most promising candidate. However, the phenomenology of the
weakly-coupled heterotic string suffers from a few long-standing problems, 
such as the stabilization of dilaton/moduli, supersymmetry breaking (by 
hidden-sector gaugino condensation), coupling unification, the strong CP 
problem, and cosmological problems involving the dilaton/moduli and 
the axion \cite{Quevedo}.
The recent developments of string duality indicate that other superstring 
theories, including M-theory, are of equal phenomenological importance. On the
other hand, string duality itself also implies that some of the problems 
associated with the weakly-coupled heterotic string theory will probably 
re-appear in the other perturbative limits (i.e., other weakly-coupled 
theories) \cite{stringduality}. Right now, it is still unclear how these 
problems will be solved eventually; one might hope to find a cure in a truly 
non-perturbative theory. This seems to be a very interesting possibility; 
however, currently our understanding of such a theory is still limited. 
Therefore, it is worth studying whether these notorious problems can be solved
and how they will be solved in the weakly-coupled heterotic string theory. 
Here, we would like to see how far one can go in this direction. We study 
these problems of the weakly-coupled heterotic string theory by adopting the 
point of view that they arise mostly due to our limited calculational power, 
little knowledge of the full vacuum structure, and an inappropriate treatment 
of gaugino condensation. As we shall see, after a more complete and consistent 
treatment, these problems could be solved or are much less severe. 

There are three major new ingredients in our treatment. The first new ingredient
is the linear multiplet formalism of the heterotic string effective theory,
where the dilaton superfield is represented by a linear supermultiplet $L$
\cite{Linear}. It was first pointed out in \cite{Gates} that the field-theoretical 
limit of the weakly-coupled heterotic string theory should be described by the 
linear multiplet formalism rather than the chiral multiplet formalism (where
the dilaton is represented by a chiral superfield.) On the other hand, there 
exists a chiral-linear duality between these two formalisms \cite{Burgess95}, 
and therefore in principle these two formalisms should be equivalent. However, 
the chiral-linear duality is apt to be very complicated, especially when full
quantum corrections are included. Therefore, there should exist a formalism
where the physics allows a simpler description. It has been argued in
\cite{Adamietz93,Derendinger94,dilaton} that, according to the above 
consideration, the linear multiplet formalism should be the more appropriate
formalism.        

The second new ingredient is a stringy non-perturbative effect. Our study of 
superstring phenomenology contains two kinds of non-perturbative effects: the 
stringy non-perturbative effects generated above the string scale, and the 
field-theoretical non-perturbative effects of gaugino condensation generated 
by strongly-interacting gauge groups below the string scale. The existence of 
significant stringy non-perturbative effects was first conjectured by S.H. Shenker 
\cite{Shenker90}. String duality and D-branes provide further evidence 
\cite{Shenker96} for Shenker's conjecture. It was first noticed by T. Banks and 
M. Dine that significant stringy non-perturbative effects could have interesting 
implications \cite{Banks94}. Here we discuss the phenomenological implications 
of stringy non-perturbative effects using the linear multiplet formalism. It is
interesting to note that, in the presence of stringy non-perturbative effects
$f(L)$ \cite{dilaton}, the coupling of heterotic string effective field theory is 
$\langle\,L/\(1+f(L)\)\,\rangle$ rather than $\langle\,L\,\rangle$ \cite{yy,multiple}. 
It is then argued that stringy non-perturbative effects are best described by
the linear multiplet formalism \cite{yy,rev}. This advantage of the linear multiplet
formalism is crucial to our study where both stringy and field-theoretical 
non-perturbative effects are considered. 

Thirdly, in the past gaugino condensate has always been described by an 
{\em unconstrained} chiral superfield $U$ corresponding to the bound state of 
$\,{\cal W}^{\alpha}{\cal W}_{\alpha}\,$ in the underlying theory. It was 
pointed out recently that $U$ should be a {\em constrained} chiral superfield 
\cite{Burgess95,Binetruy95,sduality} due to the constrained superspace 
geometry of the underlying Yang-Mills theory:
\begin{eqnarray}
U\,&=&\,-(\DbDb-8R)V, \nonumber \nonumber \\ 
\bar{U}\,&=&\,-(\DaDa-8R^{\dagger})V,
\end{eqnarray}
where $V$ is an unconstrained vector superfield. This constraint emerges from
the linear multiplet formalism naturally, and has several non-trivial 
implications \cite{Burgess95,multiple,Binetruy95}. Finally, full modular 
invariance, a very important symmetry of closed string theory \cite{ns}, is 
always maintained in our construction. It has important predictions 
\cite{rev,review}, which can be obtained only after the above ingredients of
heterotic string theory are fully taken into account. 

Based on this treatment, a simple E$_8$$\times$E$_8$ model was first studied in 
\cite{dilaton}, and a generic orbifold model was studied in \cite{multiple}.
In \cite{yy}, the analysis of \cite{dilaton,multiple} was shown to be valid in 
a more generic context. A detailed phenomenological discussion was given in 
\cite{review}, and \cite{rev} is a complete review. Due to limited space, here 
we briefly discuss several interesting phenomenological issues only.

\section{Dilaton Stabilization}
The weakly-coupled heterotic string phenomenology based on gaugino condensation 
has been long plagued by the infamous dilaton runaway problem \cite{Banks94,Dine85}, 
and there were claims in favor of the strongly-coupled heterotic string theory 
by arguing that it is unlikely that the weakly-coupled heterotic string theory 
can avoid the dilaton runaway. However, string duality implies that the strong 
coupling limit of heterotic string theory, another weakly-coupled theory (i.e., 
M-theory compactified on R$^{10}$$\times$S$^{1}$/Z$_{2}$ \cite{hw}), is plagued 
by a similar runaway problem\footnote{For example, one has to worry about the 
runaway of the interval, $\rho_{11}$, along the 11th dimension. In particular, 
$\rho_{11}$ controls supersymmetry breaking, and supersymmetry is unbroken as 
$\rho_{11}\rightarrow\infty$.} \cite{stringduality}. It was first suggested by 
T. Banks and M. Dine \cite{Banks94} that significant stringy non-perturbative 
effects could stabilize the dilaton. This proposal was studied in \cite{dilaton}  
and \cite{Casas96}\footnote{However, \cite{Casas96} did not take into account 
other aforementioned ingredients of weakly-coupled heterotic string.}. Indeed,
the dilaton can be stabilized by significant stringy non-perturbative effects;
"significant" means that stringy non-perturbative effects, $f(L)$ (in the K\"ahler
potential), satisfy the condition \cite{dilaton}:
\begin{equation}
f-L\frac{\mbox{d}f}{\mbox{d}L}\;\geq\; 2\;\;\;\;\;\;\mbox{for}\;\;\;
L\,\geq\,\CO(1). \end{equation}
This condition is actually very generic \cite{yy,multiple}, and, with reasonable
guess of $f(L)$, the dilaton is stabilized in a weak coupling regime \cite{review}.
As expected, an unsatisfactory feature is that the vanishing of cosmological
constant needs fine tuning (of $f(L)$); however, this is still an improvement
in comparison with the racetrack model \cite{rt1,rt2}\footnote{Dilaton 
stabilization in the racetrack model requires a delicate cancellation between 
contributions from different gaugino condensates, which is not natural. 
Furthermore, it has a large and negative cosmological constant when supersymmetry 
is broken.}. We emphasize that many aspects of our study are different from 
those of the racetrack model \cite{multiple,review}. For example, dilaton 
stabilization and supersymmetry breaking are possible for simple as well as
for product non-Abelian gauge groups in our study \cite{multiple}. 

\section{Moduli Physics}
\subsection{Stabilization at the Self-Dual Point}
Our study of a generic orbifold model \cite{multiple}
shows that, along with dilaton stabilization, the compactification moduli, 
$T^I$, are stabilized at the self-dual point, $\langle\,T^{I}\,\rangle=1$. 
What's more interesting is the fact that, in the vacuum (i.e., at the self-dual 
point), the $F$ components of $T^I$ vanish. Therefore, although $T^I$'s are 
stabilized by SUSY breaking effects, $T^I$'s do not contribute to the breaking 
of SUSY. Only the dilaton contributes to SUSY breaking, which leads to the 
famous dilaton-dominated scenario for soft SUSY breaking. As explained in 
\cite{multiple}, we emphasize that this unique prediction does not necessarily 
follow from any framework with modular invariance; in the weakly-coupled 
heterotic string theory this prediction is the consequence of both modular invariance 
and an appropriate treatment of gaugino condensation \cite{multiple}\footnote{This 
may explain why this prediction is absent in those works \cite{rt2} where 
modular invariance is correctly incorporated but the constraint, Eq.(1), was 
not included.}. Therefore, the weakly-coupled heterotic string theory offers a 
rationale for the well-known dilaton-dominated scenario elegantly, and a search
for the dilaton-dominated scenario might serve as a test of modular invariance
in string theory.
\subsection{Mass Hierarchy between Moduli and Gravitino} 
According to the standard lore of string phenomenology, a naive oder-of-magnitude 
estimate concludes that the dilaton and moduli have masses of order (or no larger 
than) the gravitino mass \cite{Carlos93}. These light fields with couplings 
suppressed by the Planck scale lead to the so-called cosmological moduli 
problem \cite{Carlos93,Randall94,Berkooz94}. In order to solve the cosmological 
moduli problem, there have been attempts at a hierarchy between moduli and 
squark masses \cite{Berkooz94,Nir94}; however, none of them is realistic. 
There are also possible cosmological solutions to the cosmological moduli 
problem, such as a weak scale inflation \cite{Randall94}.

It turns out that the usual estimate of dilaton and moduli masses is too rough.
In our study, the actual calculation of these masses shows that 
$m_{t^{I}}\,\approx\,(2b/b_{+})m_{\tilde G}$ \cite{multiple,review}, where 
$m_{t^{I}}$ is the moduli mass and $m_{\tilde G}$ the gravitino 
mass.\footnote{$b$ is the E$_8$ $\beta$-function coefficient, and $b_{+}$ is 
the $\beta$-function coefficient of the (largest if multi-condensation) gaugino 
condensate.}$^{,}$\footnote{The dilaton mass $m_{d}\,\geq\,m_{t^{I}}$ in general 
\cite{rev}.} For a realistic scale of gaugino condensation, $b/b_+\approx 10$ 
is required for the string models under consideration. Therefore, in contrast 
to the standard lore, there exists a natural hierarchy between the dilaton/moduli 
and squark/slepton masses, $m_{t^{I}}\approx 20 m_{\tilde G}$. This mass hierarchy  
could be sufficient to solve the cosmological moduli problem. It may have 
other non-trivial cosmological implications. Its implication on the primordial 
black hole constraints has recently been studied in \cite{Liddle}.

\section{Axion Physics}
\subsection{The Strong CP problem}The invisible axion is an elegant solution 
to the strong CP problem. However, it has been argued that QCD cannot be the 
dominant contribution to the potential of any string axion \cite{Banks96}. For 
the model-independent axion, it has been argued (using the chiral multiplet 
formalism) that the model-independent axion cannot be the QCD axion due to 
stringy non-perturbative effects of order $e^{-c\sqrt{S}}$ in the 
superpotential\footnote{$S$ is the dilaton chiral superfield.} 
\cite{Banks94,Banks96}. On the other hand, for the linear multiplet formalism 
where the dilaton is represented by a linear multiplet $L$, it is simply 
impossible to write down any $L$-dependent contribution (e.g., $e^{-c/\sqrt{L}}$) to the 
superpotential -- a constraint from holomorphy. Therefore, in our study the 
QCD axion problem of T. Banks and M. Dine \cite{Banks96} is naturally 
resolved.  

In our study of a generic orbifold model with a simple non-Abelian hidden-sector 
gauge group \cite{multiple}, the model-independent axion remains massless, and 
has the right features to be the QCD axion. As for a non-Abelian product gauge 
group which leads to multiple gaugino and matter condensation, the model-independent 
axion acquires a mass typically exponentially suppressed relative to the 
gravitino mass by a small factor of order $\langle\,\rho_2 /\rho_1\,\rangle^{1/2}$, 
where $\rho_1$ ($\rho_2$) is the gaugino condensate with the largest (second 
largest) $\beta$-function coefficient \cite{multiple}\footnote{Higher-dimension 
operators can give extra contributions to the mass of this axion. However, 
these contributions may be argued to be negligible using discrete R symmetry 
\cite{Banks94}.}. If the gauge group G$_2$ (of $\rho_2$) is reasonably smaller
than G$_1$ (of $\rho_1$), then the axion mass can still be small enough to solve
the strong CP problem\footnote{Unlike the racetrack model, in our study a delicate 
cancellation between the condensates of G$_1$ and G$_2$ is {\it not} required. 
Successful models can be constructed for the single-condensate case as well
as the multi-condensate case with G$_2$ reasonably smaller than G$_1$.}.

\subsection{Solving A Cosmological Problem}
For any of the string axions to solve the strong CP problem, there is a
cosmological constraint. The decay constant $F_{a}$ of the invisible axion 
should lie between $10^{10}$ GeV and $10^{12}$ GeV (the axion window \cite{Abbott83}). 
The upper bound, $F_{a}\leq 10^{12}$ GeV, is due to the requirement that the 
energy density of the coherent oscillations of the axion be less than the 
critical density of the universe. However, in superstring theory the axion decay 
constant $F_{a}$ is naturally of order the Planck scale, and therefore this upper 
bound is seriously violated. Although it was shown in \cite{Kim} that $F_{a}$ 
of the model-independent axion for the weakly-coupled heterotic string actually 
is $\,\approx\,10^{16}$ GeV, this is still much larger than this upper bound. 
However, cosmological constraints can be quite scheme-dependent; for example,
entropy production produced by the decays of massive particles can dilute the 
axion density and therefore raise this upper bound \cite{Dine83}. Based on the 
above idea, \cite{Moroi} proposed a refined scenario where Polonyi fields with
masses larger than about 10 TeV (in order to keep successful primordial 
nucleosynthesis) are natural candidates for the entropy production, and the 
model-independent axion is almost consistent with the new upper bound on $F_{a}$. 
Although these Polonyi fields with masses $\geq 10$ TeV seem un-natural according 
to the standard lore, this scenario of \cite{Moroi} does naturally occur in our 
study, where moduli fields ($m_{t^{I}}\approx 20 m_{\tilde G}\approx 20$ TeV)
serve the purpose of raising the upper bound on $F_{a}$ to a value consistent 
with the model-independent axion.\footnote{Note that, in our study \cite{review}, 
due to stringy non-perturbative effects the $F_a$ of model-independent axion is 
smaller than the $F_{a}\approx 10^{16}$ GeV obtained in \cite{Kim} by a factor of 
$1/50$. Therefore, in our study the value of $F_a$ is well below the new upper bound 
\cite{Moroi}.} 

\section{Newton's Constant}
It is often stated that one can determine from the low-energy values of gauge 
couplings the precise value of the gauge coupling unification scale, $M_{GUT}$, 
to be the $M_{GUT}^{(MSSM)}\,=\, 3 \times 10^{16}$ GeV based on the MSSM. 
This is a misleading statement since most string models constructed so far 
that hold a claim for being realistic include new forms of matter which 
perturb the evolution of the gauge couplings at some intermediate threshold 
\cite{unif}. In fact, as for string models considered in our study, the 
unification scale $M_{GUT}$ should naturally be the string scale $M_{s}$ 
\cite{Gaillard92}. Furthermore, the compactification scale $M_{comp}$ is also 
close to the string scale because the compactification moduli are stabilized 
at the self-dual point, $\langle\,T^{I}\,\rangle=1$. Therefore, one naturally 
expects $M_{GUT}\;\sim\;M_{s}\;\sim\;M_{comp}$.

Let's make a short remark on the Newton's constant $G_{N}$. For the 
weakly-coupled heterotic string theory, it has been shown by E. Witten 
\cite{Witten96} that there exists a lower bound on the Newton's constant:
\begin{equation}
G_{N}\;\;\geq\;\;\frac{\alpha_{GUT}^{4/3}}{M_{comp}^{2}}.
\end{equation}
If one simply takes $M_{comp}$ to be $M_{GUT}^{(MSSM)}$, the resulting lower
bound on the Newton's constant is indeed too large. On the other hand, in our
study the compactification moduli are actually stabilized at the self-dual 
point, $\langle\,T^{I}\,\rangle\,=\,1$. Therefore, the compactification scale 
is quite close to the string scale. According to the previous discussion, one 
should take $M_{comp}$ to be of order $M_{s}$, and the resulting lower bound on the 
Newton's constant is of order $\alpha_{GUT}^{4/3}/M_{s}^{2}$. This lower bound
is certainly small enough \cite{rev}. 

\section{Soft Supersymmetry Breaking Parameters}
In contrast to the studies of moduli and axion, the analysis of soft
supersymmetry breaking parameters is much more sensitive to the very details 
of a string model. A detailed discussion can be found in \cite{rev,review}. 
Here we only discuss an issue about the dilaton-dominated scenario. As explained
in Section 3.1, $\langle\,T^{I}\,\rangle=1$ and the vanishing of their 
$\langle\,F\,\rangle$ components are non-trivial results of taking into account
the aforementioned ingredients of weakly-coupled heterotic string theory, and
they lead to the well-known dilaton-dominated scenario. It is generally believed 
that a dilaton-dominated scenario results in universal soft SUSY breaking 
parameters at a high energy scale due to the universality of dilaton couplings 
\cite{FCNC}, which is a potential advantage for the FCNC constraints. However,
we would like to point out some uncertainty about this statement by studying 
the soft scalar masses for a generic orbifold \cite{multiple}. The K\"ahler
potential $K$ and the Green-Schwarz counterterm $V_{GS}$ are
\begin{equation}
K = k(V) + \sum_Ig^I +\sum_Ae^{\sum_Iq^A_Ig^I}|\Phi^A|^2 + \CO(|\Phi^A|^4),
\end{equation}
\begin{equation}
V_{GS} = b\sum_Ig^I +\sum_Ap_Ae^{\sum_Iq^A_Ig^I}|\Phi^A|^2+\CO(|\Phi^A|^4), 
\end{equation}
where $g^I=-\ln(T^I+\bar{T}^I)$, $k(V)$ is the K\"ahler potential for the 
modified linear multiplet $V$ \cite{dilaton}, $\Phi^A$'s are gauge nonsinglet 
chiral superfields and $q^I_A$'s are their modular weights. Note that $V_{GS}$,
to our knowledge, is uncertain up to modular invariant corrections in $\Phi^A$ 
(parametrized by $p_A$'s). According to \cite{review}, the scalar masses 
are:\footnote{If string threshold corrections are determined by a holomorphic 
function, they cannot contribute to the scalar masses.} 
\begin{equation}
m_A\approx\frac{\left| 1-p_A/b_{+}\right|}{1+p_A\langle\,\ell\,\rangle}m_{\tilde G}.
\end{equation}
As expected, $m_A$ does not depend on $q^I_A$ due to $\langle\,T^{I}\,\rangle=1$ 
and the vanishing of their $\langle\,F\,\rangle$ components. It is clear that 
$m_A$'s are universal -- and unwanted FCNC is thereby suppressed -- if $p_{A}$'s 
are universal \cite{FCNC}. Unfortunately, so far there is little knowledge of 
$p_{A}$'s. In general, $m_A$ is sensitive to the -- as yet unknown -- details 
of $\Phi^A$-dependent corrections to $V_{GS}$. These corrections have not been 
considered by the analyses of dilaton-dominated scenario in the past \cite{FCNC}, 
and the possibility of non-universal $p_A$'s can lead to non-universal soft 
SUSY breaking parameters even for the dilaton-dominated scenario. The 
phenomenology of several possible choices for $p_A$'s was discussed in 
\cite{review}.

\section{Concluding Remarks}
As expected, the origin of the cosmological constant remains a mystery here
although it is indeed under better control in our treatment. Again, a final 
resolution of this problem might have to wait for a complete understanding of 
superstring dynamics. The other unsettled issue is the soft SUSY breaking 
pattern. Although our study always predicts a dilaton-dominated scenario, 
in contrast to the standard lore of string phenomenology we point out that 
whether a dilaton-dominated scenario predicts universal soft SUSY breaking 
parameters actually depends on whether the matter couplings to the 
Green-Schwarz counterterm are universal. To settle this issue, a better 
understanding of the matter dependence of the Green-Schwarz counterterm for 
generic string models is certainly required; it deserves further studies and 
could lead to a rich phenomenology. In conclusion, we emphasize that this work 
is certainly not final, and it is very important to understand more about the 
non-perturbative aspects of superstring theories, realistic string model 
building and the phenomenology. After a careful re-examination of the problems 
of the weakly-coupled heterotic string theory, it is also hoped that confusion 
about the current status of weakly-coupled heterotic string theory is 
clarified by this work.
\section*{Acknowledgements}
\hspace{0.8cm}
This short review is based on my work \cite{dilaton,multiple,review} in 
collaboration with Pierre Bin\'etruy and Mary K. Gaillard.
This work was supported in part by the Director, Office of 
Energy Research, Office of High Energy and Nuclear Physics, Division 
of High Energy Physics of the U.S. Department of Energy under Contract 
DE-AC03-76SF00098 and in part by the National Science Foundation under 
grant PHY-95-14797.
\pagebreak

\end{document}